\documentclass[review,3p,12pt]{elsarticle}
\usepackage{graphicx}
\usepackage{subfig}
\usepackage{color}
\usepackage{amsmath}
\usepackage{amssymb}
\usepackage{caption}
\usepackage{upgreek}
\usepackage{epstopdf}
\usepackage{xcolor}
\usepackage{pdfsync}
\usepackage{rotating}
\usepackage{textcomp}
\usepackage[rmdefault=mdput,
            OMLmathrm,
            OMLmathbf         
           ]{isomath}
\usepackage{adjustbox}
\usepackage{multirow}
\biboptions{sort&compress,square}
\journal{-}
\begin{document}
\begin{frontmatter}
\title{Transfer learning for predicting source terms of principal component transport in chemically reactive flow}
\author[SNL]{Ki Sung Jung\corref{cor1}}
\ead{kjung@sandia.gov}
\cortext[cor1]{Corresponding authors.}
\author[NCSU]{Tarek Echekki}
\author[SNL]{Jacqueline H. Chen}
\author[SNL]{Mohammad Khalil}
\address[SNL]{Sandia National Laboratories, Livermore, CA 94551-0969, USA}
\address[NCSU]{Department of Mechanical and Aerospace Engineering, North Carolina State University, Campus Box 7910, Raleigh 27695, NC, USA}
\begin{abstract}
Although data-driven reduced-order models have been recently applied to chemically reactive flows, one of the drawbacks of these models is associated with their high dependency on the training dataset. The training dataset of a reduced-order model is often comprised of multi-dimensional numerical simulations, such that the number of high-quality training samples at a given operating condition could be sparse under practical scenarios. Transfer learning has been highlighted as a promising framework to increase the accuracy of the data-driven model in the case of data sparsity, specifically by leveraging pre-trained knowledge to the training of the target model. The objective of this study is to evaluate whether the number of requisite training samples can be reduced with the use of various transfer learning models for predicting, for example,  the chemical source terms of the data-driven reduced-order model that represents the homogeneous ignition process of a hydrogen/air mixture. Principal component analysis is applied to reduce the dimensionality of the hydrogen/air mixture in composition space. Artificial neural networks (ANNs) are used to tabulate the reaction rates of principal components, and subsequently, a system of ordinary differential equations is solved. As the number of training samples decreases at the target task (i.e., for $T_0 >$ 1000 K and various $\phi$), the reduced-order model fails to predict the ignition evolution of a hydrogen/air mixture. Three transfer learning strategies are then applied to the training of the ANN model with a sparse dataset. The performance of the reduced-order model with a sparse dataset is found to be remarkably enhanced if the training of the ANN model is restricted by a regularization term that controls the degree of knowledge transfer from source to target tasks. To this end,  a novel transfer learning method is introduced, parameter control via partial initialization and regularization (PaPIR), whereby the amount of knowledge transferred is systemically adjusted for the initialization and regularization of the ANN model in the target task. It is found that an additional performance gain can be achieved by changing the initialization scheme of the ANN model in the target task when the task similarity between source and target tasks is relatively low. 
\end{abstract}
\end{frontmatter}

\section{Introduction}\label{intro}
With the continuous advancement of chemical kinetic mechanisms, detailed chemical mechanisms of large-hydrocarbon fuels can comprised of thousands of species and tens of thousands of elementary chemical reactions \cite{Lu09}. The sheer size of the mechanisms is a major challenge for high-fidelity numerical simulations of turbulent reacting flows with large-hydrocarbon fuels.  This is mainly because of the high dimensionality of the thermo-chemical state  and their wide range of temporal scales. Various approaches have been developed to reduce the number of variables in composition space. These include skeletal/reduced chemical kinetic mechanisms, in which key species and elementary chemical reactions are extracted from a detailed mechanism using techniques such as directed relation graph (DRG) \cite{Lu05}, DRG with error propagation \cite{Pepiot08}, computational singular perturbation \cite{Lam94}, and path flux analysis \cite{Sun10}, thereby reducing the overall size and computational cost of simulations.

More recently, a data-based dimensionality reduction method has also been applied to chemically reactive flows, where a low-dimensional manifold of the original thermochemical state variables is defined based on data-based dimensionality reduction techniques, including linear and non-linear principal component analysis (PCA) \cite{Sutherland09,Parente09,Mirgolbabaei13,Mirgolbabaei14}. The distinct features of the data-driven technique compared to physics-based low-dimensional manifolds, such as steady laminar flamelet model \cite{Peters83,Peters84}, unsteady flamelet/progress variable approach \cite{Pierce04, Ihme05}, and flamelet generated manifold \cite{Oijen00}, are that the correlations of the thermochemical state vector are identified by a ``training dataset'' that is prepared \textit{a priori}, and the rank of the low-dimensional manifold can be easily adjusted by the user depending on the trade-off between the compression ratio and truncation error. Either linear mapping or a non-linear regression method (e.g., artificial neural network (ANN), Gaussian process regression (GPR)) is employed for the closure of the governing equations \cite{Sutherland09,Isaac15}. A PCA-based reduced-order model (ROM) has been shown to replicate  characteristics of turbulent flames through \textit{a priori} \cite{Sutherland09,Mirgolbabaei13, Parente13, Dalakoti21} and \textit{a posteriori} evaluations \cite{Echekki15,Biglari15, Owoyele17,Malik21, Malik22,Kumar23,Abdelwahid23}. In their recent study, Kumar et al.~\cite{Kumar23} have demonstrated the potential speed up of surrogate 3D DNS involving PCA-based ROM compared to DNS involving the solution for species and energy equations for methane-air premixed flames stabilized on a slot burner.

Despite the advantages of the data-based ROM for reactive flow simulations, one of the drawbacks of the model is associated with its strong dependency on the quality of training data. For instance, Owoyele and Echekki performed two-dimensional (2-D) and three-dimensional (3-D) surrogate direct numerical simulations (DNS) of a premixed methane/air flame in a vortical flow with the transport of principal components \cite{Owoyele17}, revealing that a low-dimensional manifold defined from a one-dimensional (1-D) training dataset fails to reproduce 2-D flame characteristics due to a lack of information on the curvature effect in the 1-D training dataset. Dalakoti et al. \cite{Dalakoti21} also pointed out that a PCA-based ROM based on  either a zero-dimensional (0-D) homogeneous reactor or a 1-D non-premixed igniting flamelet dataset is unable to fully represent the heat release characteristics of a 3-D spatially-developing turbulent \textit{n}-dodecane jet flame at high-pressure conditions. These findings indicate that a data-driven ROM for chemically reactive flows  requires high-quality training data, usually obtained by carrying out multi-dimensional simulations with a detailed chemical kinetic mechanism, to reproduce the characteristics of the full-order model (FOM) accurately. However, given that one of the main purposes of adopting ROMs for reactive flow simulations is to alleviate computational cost, an argument can be made that it would be impractical to always obtain a sufficient number of high-quality training samples whenever operating conditions of a combustion system change. In other words, the amount of high-quality training data, necessary to optimize a ROM for chemically reactive flows with limited computational resources,  can be sparse under practical conditions. 


In the machine learning community, transfer learning has been highlighted as a promising framework to improve the performance in the case of data sparsity, together with providing a robust initialization scheme and speeding up the learning process \cite{Pan10}. The central idea of transfer learning in the context of machine learning is that a pre-trained machine learning model, optimized with a sufficient number of training samples, is utilized to train a target machine learning model that has sparse training data. Numerous studies demonstrate that the performance of the machine learning model with a sparse dataset is remarkably enhanced by applying transfer learning for clustering \cite{Yang09,Mieth19,Wang21}, classification \cite{Quattoni08,Yao10,Zhu11,Hosny18}, and regression cases \cite{Salaken19,Subel21,Yang22,Liu22}. There are different ways of ``transferring'' knowledge from the previous model (or source model) to the target model, such as instance-based algorithms, feature-based algorithms, model-based algorithms, and relation-based algorithms \cite{Yang20}. In the present study, a model-based transfer learning algorithm, also known as parameter-based transfer learning \cite{Yang20}, is adopted to utilize the parameters obtained from the previous machine learning model for the optimization of the target model with sparse training samples. The straightforward way is to freeze all (or some of) the parameters of the target machine learning model with those obtained from the pre-trained model \cite{Pan10}. The parameters of the previous machine learning model can also be used as an initial guess of the parameter values in the target machine learning model. A regularization-based transfer learning method has recently been introduced \cite{Li18, Li20, De22} in which the knowledge of the previous machine learning model can be ``partially'' transferred to the target model by adjusting the magnitude of the regularization parameter.

The main objective of the present study is to investigate a possibility of alleviating the requisite number of training samples for optimizing data driven ROM for chemically reactive flows by utilizing different transfer learning methods. It has been revealed that an accurate prediction of the source term is one of the most challenging parts within the framework of data-driven ROM \cite{Dalakoti21}. Therefore, the main focus of the present study is to utilize transfer learning methods to mitigate the requisite number of training samples in the prediction of a 0-D homogeneous ignition process for a hydrogen/air mixture in a constant volume reactor. While the PC-transport ROM has non-stiff transport terms \cite{Sutherland09}, these terms can be easily predicted by using a shallow neural network model compared to the source term \cite{Owoyele17,Kumar23}. 

The dimensionality of the hydrogen/air mixture in composition space is reduced by applying PCA, and the non-linear relationship between the principal components and their reaction rates is tabulated by optimizing ANN models. The effect of the number of training samples on the performance of the data-driven ROM is first investigated, and subsequently, different transfer learning approaches are adopted to predict the reaction rates of the principal components with a sparse dataset. To this end, we introduce a novel transfer learning method called ``Parameter control via Partial Initialization and Regularization (PaPIR)'', where the amount of knowledge transferred from source to target ANN model can be systemically adjusted for the initialization and regularization of the target ANN model. 

The outline of the paper is as follows. Section~\ref{Method} presents the details of the data-driven ROM, ANN models, and various transfer learning methods. Section~\ref{source_task} illustrates the results of the PCA-based data-driven ROM for the 0-D ignition process of a hydrogen/air mixture with various initial conditions depending on the number of training samples. Transfer learning is not applied thus far to highlight the importance of the training data on the performance of the model. In Section~\ref{TL}, four different transfer learning methods are utilized for the training of the ANN model with a sparse dataset, from which the performance of the transfer learning methods for various target tasks is evaluated. 

\section{Methodology}\label{Method}
Homogeneous ignition of a hydrogen/air mixture in a constant volume reactor is predicted by applying  PCA-based data-driven reduced-order model (PC-transport ROM). Since integration of numerically stiff chemistry is a bottleneck for many reactive flow simulations, it is reasonable to consider that the present  homogeneous reactor configuration is an important benchmark case for evaluating the ability of PC-transport ROM to accurately reproduce reactive flow simulations. 

The temporal evolution of the original thermochemical-state vector with different initial conditions is first collected by performing a series of 0-D simulations of the homogeneous hydrogen/air mixture, and subsequently, a low-dimensional manifold is defined by applying PCA to the collected data. Here, the new variables defined by PCA are denoted as the principal components (PCs). The reaction rates of the PCs are tabulated as a function of PCs using an ANN. After training the ANN model on one task, the knowledge of the trained ANN model is transferred to another task where the size of the training data is assumed to be sparse. The performance of the different transfer learning methods is then systemically investigated by varying (1) the task similarity between the source and target tasks, and (2) the degree of data sparsity in the target task. The methodology of these investigations is described in this section. 

\subsection{0-D ignition dataset for a homogeneous hydrogen/air mixture}\label{0-D-Data}

In a spatially-homogeneous constant volume reactor, the temporal evolution of species and temperature starting from the initial time, $t$ = 0, is computed by solving the system of ordinary differential equations (ODEs) defined by

\begin{equation}
  \frac{d \boldsymbol{\mathrm{\theta}}}{dt} = \dot{\boldsymbol{\mathrm{\omega}}}_{\boldsymbol{\mathrm{\theta}}}, t \in \left[0, t_{f}\right]
  \label{0-D}
\end{equation}
where $\boldsymbol{\mathrm{\theta}}$ represents the thermochemical state vector (i.e., species mass fraction and temperature),  $\dot{\boldsymbol{\mathrm{\omega}}}_{\boldsymbol{\mathrm{\theta}}}$ the reaction rate vector of $\boldsymbol{\mathrm{\theta}}$, and $t_{f}$ the end time. For the initial conditions, the initial pressure of the system, $p_0$, is fixed to be atmospheric, and different values for the initial temperature, $T_0$, are used including 1000, 1050, 1100, 1300, and 1400 K. The initial mass fractions of the hydrogen/air mixture are determined by an equivalence ratio, $\phi$, which ranges from 0.1 to 3.0. A detailed chemical kinetic mechanism for hydrogen/air mixtures, developed by Li et al. \cite{Li04}, is used where the dimension of the original thermochemical state vector is 10. A six-stage, fourth-order Runge-Kutta method \cite{Kennedy94} with a uniform time step, \textit{dt}, of 0.2 ns is adopted for time integration. A CHEMKIN library \cite{Kee96} is used to compute the chemical kinetics and thermodynamic properties of the mixture.

\begin{figure}
  \centerline{\includegraphics[width=70mm]{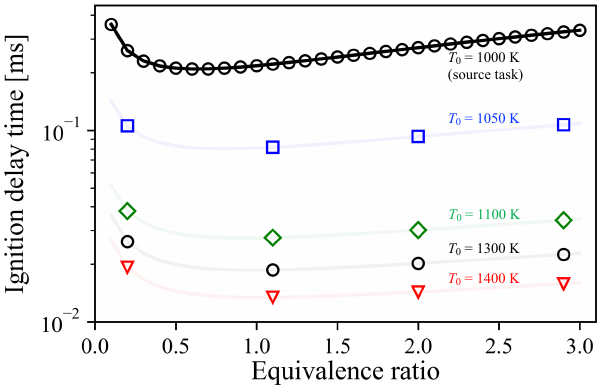}}
  \caption{Variations in 0-D ignition delay time, $\tau_{\rm{ig}}$, of the hydrogen/air mixture for different initial temperature, $T_0$, as a function of equivalence ratio, $\phi$. In the present study, it is assumed that the number of training samples at the source task ($T_0$ = 1000 K) is sufficient, while the number of training samples at the target tasks ($T_0 > $ 1000 K) is sparse.}\label{Fig1}
\end{figure}

Figure~\ref{Fig1} shows the ignition delay time, $\tau_{\rm{ig}}$, of the hydrogen/air mixture for various $\phi$ and $T_0$. As expected, the variation in $\tau_{\rm{ig}}$ exhibits a ``U''-shaped profile as a function of $\phi$. Here, $\tau_{\rm{ig}}$ is defined as the time at which the temperature gradient is maximum. In accordance with Arrhenius Law, $\tau_{\rm{ig}}$ notably changes with changes in $T_0$. 

In the present study, the objective of the PC-transport ROM is to replicate the ignition characteristics of the hydrogen/air mixture over a wide range of $\phi$ at a specific $T_0$, meaning that training samples, the low-dimensional manifold, and the corresponding ANN model are separated by $T_0$. To reasonably provide a data-sparse scenario, an underlying assumption of the present study is that there exists a sufficient number of training samples spanning over $\phi$ at $T_0$ of 1000 K, while the  training data size for the cases where $T_0 >$ 1000 K is assumed to be sparse (see the symbols in Fig.~\ref{Fig1} as an example). Specifically, at $T_0 =$ 1000 K, the training dataset is collected by carrying out 30 different 0-D simulations varying $\phi$ (i.e., $\Delta \phi$ = 0.1; $\phi$ ranging from 0.1 to 3.0), and then the low-dimensional manifold is defined by applying PCA to the training dataset. Training of the ANN model by using a sufficient number of training samples at $T_0 =$ 1000 K is considered as the ``source task'' for the present study. Here, $N_{\phi}$ is defined as the number of 0-D simulations at a given $T_0$ such that $N_{\phi}$ of the source task is 30. For the ``target tasks'' where $T_0$ is higher than 1000 K (i.e., 1050, 1100 1300, and 1400 K), $N_{\phi}$ is set to be less than or equal to 4, such that the number of training data for the target task is forced to be sparse. In this study, a ``sparse dataset'' refers to a dataset with insufficient training samples such that the corresponding ROM is unable to replicate the ignition characteristics of a fuel/air mixture with a wide range of $\phi$ (i.e., $\phi = 0.1 - 3.0$) at a given $T_0$. The description of the dataset with different $N_{\phi}$ is summarized in Table~\ref{Table1}. 

Note that the present study provides a data-sparse scenario based on a series of 0-D simulations, but such an imbalance in the number of training samples can also be observed from multi-dimensional simulations and experiments \cite{Humbird20,Subel21}. It is also noted that transfer learning methods used in the present study are not limited to specific source and target tasks. Rather, these methods can be applied to various scenarios (e.g., different pressure or equivalence ratio conditions between source and target tasks), provided that there is a task similarity between two tasks. 

For each 0-D simulation, the thermochemical state vector and their reaction rate, $\boldsymbol{\mathrm{\theta}}$ and $\dot{\boldsymbol{\mathrm{\omega}}}_{\boldsymbol{\mathrm{\theta}}}$, respectively, are uniformly sampled from $t$ of 0 to 2$\tau_{\rm{ig}}$. The number of samples for each 0-D simulation is set to be 20,000 such that the first 10,000 samples are assigned to the pre-ignition zone and the remaining 10,000 samples are related to the post-ignition zone. The ``test dataset'' at a given $T_0$ is also prepared to evaluate the accuracy of the PC-transport ROM. It consists of 29 different 0-D simulation results at a given $T_0$ ($\Delta \phi$ = 0.1; $\phi$ ranging from 0.15 to 2.95) and is separated from the training dataset. 

\begin{table}
  \footnotesize
  \caption{Description of the dataset with different $N_{\phi}$.}
  \begin{minipage}{\linewidth}
    \begin{center}
    \begin{tabular}{c|c|c}
      \hline    
      $N_{\phi}$   &  $\phi$ & Number of samples (\textit{M})\\ 
      \hline
      2 & 0.5, 1.5  & 40,000 \\
      \hline      
      3 & 0.5 ,1.5, 2.5  & 60,000\\
      \hline      
      4 & 0.2, 1.0, 2.0, 3.0  & 80,000\\
      \hline      
      30 & 0.1 $\sim$ 3.0 ($\Delta \phi$ = 0.1) & 600,000\\
      \hline      
    \end{tabular}
    \end{center}
  \end{minipage}
  \label{Table1}
\end{table}

\subsection{Principal component analysis}\label{PC}
Consistent with previous studies applying PC-transport ROM \cite{Sutherland09,Parente09,Mirgolbabaei13,Mirgolbabaei14}, the dimension of the original thermochemical vector is reduced by applying PCA. Assuming that $M$ number of samples of the $N-$dimensional thermochemical state vectors are collected by performing multiple 0-D simulations at a given $T_0$, the dataset of the thermochemical vector, $\boldsymbol{\mathrm{\Theta}}$ = [$\boldsymbol{\mathrm{\theta}}_1$, $\boldsymbol{\mathrm{\theta}}_2$, ..., $\boldsymbol{\mathrm{\theta}}_M$], is composed of a $N \times M$ dimensional matrix. After normalizing $\boldsymbol{\mathrm{\Theta}}$ based on its component-wise range, the $N \times N$ dimensional matrix of orthonormal eigenvectors, \textbf{Q}$^T$, of the covariance matrix of $\boldsymbol{\mathrm{\Theta}}$ is constructed, and subsequently, the dataset of the PC vector, $\boldsymbol{\mathrm{\Psi}}$, can be defined as

\begin{equation}
  \boldsymbol{\mathrm{\Psi}} = \textbf{Q}^{\rm{T}} \boldsymbol{\mathrm{\Theta}}
  \label{PC-transport}
\end{equation}
where $\boldsymbol{\mathrm{\Psi}} \in \mathbb{R}^{N \times M}$ represents the $M$ numbers of collections of the PC vector, $\boldsymbol{\mathrm{\psi}} =$ [$\psi_1$, $\psi_2$, ..., $\psi_{N}$]$^{\rm{T}}$. 

Note that the first PC, $\psi_1$, is a linear combination of the original thermochemical state vector that captures the maximum variance of the dataset. The second PC, $\psi_2$, is then orthogonal to the first PC, and all the subsequent PCs follow the same concept. In the present study, the leading first five PCs (i.e., $N_{\rm{PC}}$ = 5) are retained from $\boldsymbol{\mathrm{\psi}}$ such that the dimensionality of the system is reduced from 10 to 5, which captures over 99\% of the original total variance. In other words, a $N \times N_{\rm{PC}}$ matrix of $\textbf{A}$ is constructed that contains the leading $N_{\rm{PC}}$ eigenvectors of $\textbf{Q}$. The low-dimensional manifold then becomes

\begin{equation}
  \boldsymbol{\mathrm{\Psi}}^{red} = \textbf{A}^{\rm{T}} \boldsymbol{\mathrm{\Theta}}
  \label{PC-transport2}
\end{equation}
where $\boldsymbol{\mathrm{\Psi}}^{red} \in \mathbb{R}^{N_{\rm{PC}} \times M}$ represents the dataset of the truncated PC vector, $(\boldsymbol{\mathrm{\psi}}^{red})$ = [$\psi_1$, $\psi_2$, ..., $\psi_{N_{\rm{PC}}}$]$^{\rm{T}}$. Hereinafter, $\boldsymbol{\mathrm{\Psi}}^{red}$ and $\boldsymbol{\mathrm{\psi}}^{red}$ are referred to as $\boldsymbol{\mathrm{\Psi}}$ and $\boldsymbol{\mathrm{\psi}}$, respectively, for the sake of brevity. 

The system of ODEs for the low-dimensional manifold can be defined by projecting Eq.~\ref{0-D} on the matrix $\textbf{A}^{\rm{T}}$:

\begin{equation}
  \frac{d \boldsymbol{\mathrm{\psi}}}{dt} = \dot{\boldsymbol{\mathrm{\omega}}}_{\boldsymbol{\mathrm{\psi}}}, t \in \left[0, t_{f}\right]
  \label{PC-ODE}
\end{equation}
where $\dot{\boldsymbol{\mathrm{\omega}}}_{\boldsymbol{\mathrm{\psi}}}$ is the reaction rate term for $\boldsymbol{\mathrm{\psi}}$, defined by $\dot{\boldsymbol{\mathrm{\omega}}}_{\boldsymbol{\mathrm{\psi}}}$ = $\textbf{A}^{\rm{T}} \dot{\boldsymbol{\mathrm{\omega}}}_{\boldsymbol{\mathrm{\theta}}}$ \cite{Sutherland09}. In the framework of PC-transport ROM, the time integration of Eq.~\ref{PC-ODE} is solved instead of solving Eq.~\ref{0-D}, and then a conversion from $\boldsymbol{\mathrm{\psi}}$ into $\boldsymbol{\mathrm{\theta}}$ is carried out as a post-processing step. An ANN model is used for the tabulation of $\dot{\boldsymbol{\mathrm{\omega}}}_{\boldsymbol{\mathrm{\psi}}}$ as a function of $\boldsymbol{\mathrm{\psi}}$. 

\begin{figure}
  \centerline{\includegraphics[width=70mm]{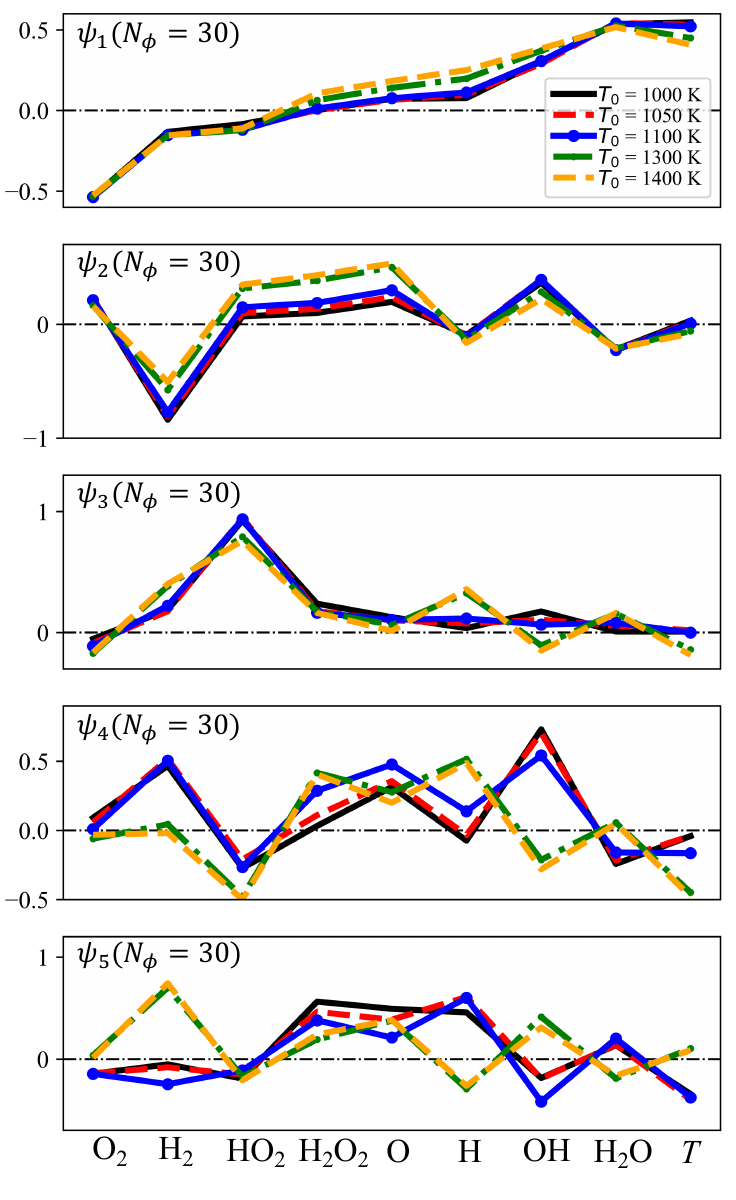}}
  \caption{Modes of the first five PCs depending on the training dataset varying $T_0$ with $N_{\phi}$ of 30.}\label{Fig2a}
\end{figure}

 
Figure~\ref{Fig2a} shows the PC modes with respect to the original thermochemical vector defined by using different training dataset in terms of $T_0$. It is readily observed from the figure that the PC modes show a similar trend irrespective of the dataset. Specifically, the first PC mode is negatively correlated with the fuel and oxidizer, while it is positively correlated with the product (i.e., H$_2$O) and temperature. Accordingly, the first PC represents the oxidation progress of the hydrogen/air mixture. The second PC mode is primarily correlated with the fuel, and the third PC mode is correlated with the formation of HO$_2$. The results indicate that the PCs obtained through the data-driven approach are linked to a physical interpretation of the combustion system, consistent with previous findings \cite{Owoyele17, Malik21}.

Nonetheless, it is important to note that the PC modes are slightly altered with change of $T_0$, which can have a significant impact on the application of transfer learning to the ROM. In other words, a unified definition of the low-dimensional manifold throughout tasks would be preferred to transfer the knowledge efficiently. In the present study, $\textbf{A}^{\rm{T}}$ defined from the source task (i.e., $N_{\phi} = 30$ and $T_0$ = 1000 K) is applied to all target tasks to ensure consistency in the definitions of $\boldsymbol{\mathrm{\psi}}$ and $\dot{\boldsymbol{\mathrm{\omega}}}_{\boldsymbol{\mathrm{\psi}}}$. Such an approach is based on the observation that despite the presence of slight differences in the PC modes, the first PC, which also accounts for most of the data variance, also exhibits the least difference when $T_0$ is varied.

Note, however, that using a unified definition of $\textbf{A}^{\rm{T}}$ has a potential risk of introducing noticeable errors during the conversion from $\boldsymbol{\mathrm{\psi}}$ to $\boldsymbol{\mathrm{\theta}}$ in the target task, especially if the reconstruction is carried out by using a matrix conversion step, $\boldsymbol{\mathrm{\theta}} \approx \textbf{A} \boldsymbol{\mathrm{\psi}}$. To address this issue, another non-linear ANN model is employed to convert from $\boldsymbol{\mathrm{\psi}}$ to $\boldsymbol{\mathrm{\theta}}$ for all cases, instead of using the matrix inversion. This ensures that the performance of the reconstruction is mainly affected by the number of samples, $M$, in the target task rather than the choice of $\textbf{A}^{\rm{T}}$. As will be discussed, such an ANN model is found to require far fewer parameters compared to the other ANN models that predict $\dot{\boldsymbol{\mathrm{\omega}}}_{\boldsymbol{\mathrm{\psi}}}$. Hence, this ANN model shows a reasonable accuracy even trained with a sparse dataset. It is also noted that the non-linear ANN was found to reduce the reconstruction error during the conversion from $\boldsymbol{\mathrm{\psi}}$ to $\boldsymbol{\mathrm{\theta}}$ as compared to the matrix inversion, $\boldsymbol{\mathrm{\theta}} \approx \textbf{A} \boldsymbol{\mathrm{\psi}}$ \cite{Mirgolbabaei15}. 

\begin{figure}
  \centerline{\includegraphics[width=70mm]{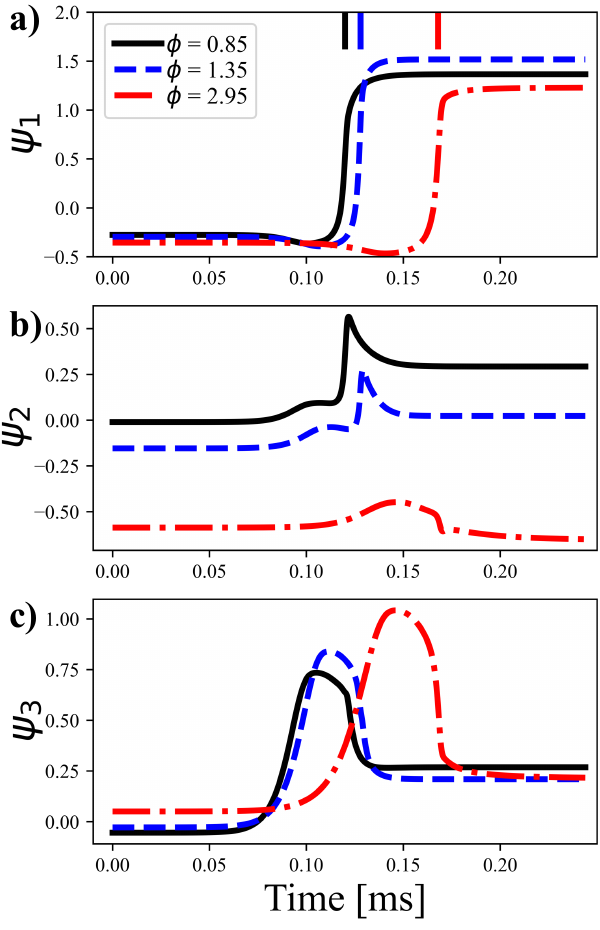}}
  \caption{Temporal evolution of the first three principal components for three different equivalence ratios, $\phi$, of 0.85, 1.35, and 2.95, obtained by projecting $\textbf{A}^{\rm{T}}$ onto the FOM result. The vertical lines in (a) represent the ignition delay time for different $\phi$. Here, the ignition delay time is defined by the time at which the temperature gradient reaches its maximum value.}\label{Fig3}
\end{figure}

Figure~\ref{Fig3} shows the temporal evolution of the first three PCs with three different $\phi$ of 0.85, 1.35, and 1.85 at $T_0$ of 1000 K, obtained by projecting $\textbf{A}^{\rm{T}}$ onto the FOM result. As discussed earlier, the first PC represents the progress variable of the mixture, such that the first PC switches from negative to positive values near the ignition delay time. For the third PC, its mode is mainly correlated with the intermediate species, namely HO$_2$, such that it is maximum just before ignition of the mixture. 

\subsection{Artificial neural network}\label{NN}

A fully connected, multi-input, and multi-output ANN model is used to predict the reaction rates of the PCs. A PC vector is used as an input of the ANN, and the reaction rates of the PCs are the output of the ANN model. The architecture of the ANN model is determined by performing a grid search method, from which the number of hidden layers and nodes are set to 3 and 30, respectively. The hyperbolic tangent activation function is adopted for all hidden layers. For cases without the application of transfer learning methods, the Xavier normal initialization method \cite{Xavier10}, which is a commonly used initialization scheme that is compatible with the hyperbolic tangent activation function, is employed.


As discussed in Section~\ref{PC}, the other ANN model is also employed and trained for the reconstruction from $\boldsymbol{\mathrm{\psi}}$ to $\boldsymbol{\mathrm{\theta}}$. One hidden layer with 10 nodes is found to be sufficient for this ANN model to reconstruct the original thermochemical scalars with reasonable accuracy. Note that such an ANN model requires  considerably fewer number of training samples, and therefore, transfer learning is not applied to this model. 



For both source and target tasks, 80\% of the dataset is used as a training dataset, and the remaining 20\% is allocated as the validation set to assess the model's performance and to prevent overfitting. Mean absolute error (MAE) loss function is applied for the ANN training, consistent with previous studies showing that MAE shows a better performance than RMSE and MSE in capturing an ignition process of a fuel/air mixture \cite{Zhang20,Han22}. The Adam optimizer \cite{Adam} is used for stochastic optimization. Once the ANN model is optimized using the training dataset, it is both \textit{a priori} and \textit{a posteriori} evaluated against the test dataset, which is not involved in the training process.

To efficiently capture the ignition process of the hydrogen/air mixture, the dataset is decomposed into three clusters (Cl\#1--Cl\#3): (Cl\#1) earlier ignition period ($\psi_{3} - \psi_{3,0} <$ 0.005 and $\psi_{1} < 0.0$), (Cl\#2) later ignition period ($\psi_{3} - \psi_{3,0} \ge $ 0.005 and $\psi_{1} < 0.0$), and (Cl\#3) post ignition period ($\psi_{1} \ge 0.0$), where the $\psi_{3,0}$ denotes the magnitude of $\psi_{3}$ at the initial condition. The clustering criteria are based on the observation that $\psi_{1}$ and $\psi_{3}$ effectively represent the progress variable and evolution of intermediate species, respectively, as depicted in Fig.~\ref{Fig3}. Note that the data clustering method has been proven as an effective way to capture the ignition process of various fuel/air mixtures \cite{Zhang20,Han22}. 

\subsection{Transfer learning methods}\label{TL-methods}

In this study, four different transfer learning methods are applied to the target tasks. Let \textbf{\textit{h}}$^s$ denote the parameter vector extracted from the pre-trained source task. The first transfer learning method (TL1) is that the knowledge of the pre-trained ANN model obtained from the source task is fully shared with the target task. In other words, the parameter vector in the target task, denoted by \textbf{\textit{h}}, is identical to \textbf{\textit{h}}$^s$, and no fine-tuning step is performed in TL1. Therefore, it can be conjectured that TL1 is likely to show good performance only when the task similarity between the source and target is very high. The second transfer learning method (TL2) is to set the initial parameter vector in the target task, \textbf{\textit{h}}$_0$, with \textbf{\textit{h}}$^s$, and then fine-tune the model using the sparse dataset in the target task. In other words, TL2 serves to initialize the ANN model in the target task by using \textbf{\textit{h}}$^s$, and as such, \textbf{\textit{h}} will be different from \textbf{\textit{h}}$^s$ after fine-tuning. 

As discussed in \cite{Li20}, a drawback of TL2 is that previous knowledge obtained from the source task may be lost during the fine-tuning step. To resolve this issue, the third transfer learning method (TL3), which is associated with parameter restriction, is applied. The total loss function, $\mathcal{L}$, in TL3 includes the regularization term \cite{Li18, Li20}, which is slightly different from the conventional $l_2$ regularizer, as follows:

\begin{equation}
  \mathcal{L} = \text{MAE} + \lambda_1 \lVert \textbf{\textit{h}} - {\textbf{\textit{h}}^{s} \rVert_2^2}
  \label{TL3}
\end{equation}
where MAE represents the mean absolute error loss function, and $\lambda_1$ is the regularization parameter. Here, the magnitude of $\lambda_1$ mainly controls the degree of knowledge transferred from source to target task during the fine-tuning step in the target task. Consistent with TL2, \textbf{\textit{h}}$_0$ in TL3 is set to be \textbf{\textit{h}}$^s$, and subsequently, the ANN model is fine-tuned by using the dataset in the target task. It is evident that $\textbf{\textit{h}}$ will be identical to \textbf{\textit{h}}$^s$ as the magnitude of $\lambda_1$ is very high, whereas there is no penalty for $\textbf{\textit{h}}$ to change during the fine-tuning step at $\lambda_1 = 0$. Therefore, it can be considered that TL3 becomes equivalent to TL1 and TL2 as the magnitude of $\lambda_1$ approaches infinity and zero, respectively. 

Lastly, we introduce a novel transfer learning method called ``Parameter control via Partial Initialization and Regularization (PaPIR)''. The central idea of PaPIR is to provide a unified transfer learning framework in terms of initialization and regularization. In addition to applying $\lambda_1$ to adjust the effect of the regularization during training, another variable, $\lambda_2$, is introduced in PaPIR so that the amount of previous knowledge transferred to the target task in terms of the initialization can also be controlled by changing the magnitude of $\lambda_2$. The initialization method of the PaPIR is a combination of two initialization schemes, namely Xavier normal initialization \cite{Xavier10} and initialization with \textbf{\textit{h}}$^{s}$. Xavier normal initialization is a family of the Gaussian-based initialization technique with zero mean and a determined variance. Thus, it is a sort of random initialization strategy and in unrelated to the pre-trained knowledge. On the other hand, initialization with \textbf{\textit{h}}$^{s}$ is categorized as a data-driven initialization strategy \cite{Narkhede22}. 

The initialization process in PaPIR follows a normal distribution function, $N$, which is expressed as follows: 

\begin{equation}
  \textbf{\textit{w}}_{0} = \textit{N} \left( \lambda_2 \textbf{\textit{w}}^{s}, \left[ \sqrt{\frac{2}{f_{i} + f_{o}}} \left( 1 - \lambda_2 \right) \right]^2 \right)
  \label{PaPIR-1}
\end{equation}

\begin{equation}
  \textbf{\textit{b}}_{0,b} = \textit{N} \left( \lambda_2 \textbf{\textit{b}}^{s}, 0^2 \right) 
  \label{PaPIR-2}
\end{equation}
where \textbf{\textit{w}}$_{0}$ and \textbf{\textit{b}}$_{0}$ represent the initial weights and biases vector of the ANN model in the target task (i.e., \textbf{\textit{h}}$_{0}$ = [\textbf{\textit{w}}$_{0}$, \textbf{\textit{b}}$_{0}$]), and \textbf{\textit{w}}$^{s}$ and \textbf{\textit{b}}$^{s}$ represent the weight and biases vector extracted from the source task (i.e., \textbf{\textit{h}}$^{s}$ = [\textbf{\textit{w}}$^{s}$, \textbf{\textit{b}}$^{s}$]). Here, $f_{\rm{i}}$ and $f_{\rm{o}}$ represent the number of incoming and outgoing nodes at each layer, respectively, which are identical to those used in the Xavier normal initialization scheme \cite{Xavier10}. 

As $\lambda_2$ in Eqs.~\ref{PaPIR-1}--\ref{PaPIR-2} approaches zero, the initialization scheme becomes equivalent to the Xavier normal initialization method. As $\lambda_2$ approaches unity, on the other hand, \textbf{\textit{h}}$_{0}$ simply becomes identical to \textbf{\textit{h}}$^{s}$. Thus, the degree of knowledge transfer for the initialization of the target task can be adjusted by varying the value of $\lambda_2$ between zero and unity, which equivalently represents a bound between the Xavier normal initialization method and \textbf{\textit{h}}$^{s}$, respectively. As will be demonstrated in Section~\ref{source_task}, PC-transport ROM with a sparse dataset generally fails to capture the overall ignition process of a hydrogen/air mixture if the ANN model is trained from scratch. This shortcoming is mainly attributed to the propensity of the ANN models to get stuck in local minima, especially with a sparse dataset. Given that an appropriate initialization scheme can help avoid local minima \cite{Narkhede22}, PaPIR has the potential advantage of enhancing the performance of transfer learning by introducing some degree of randomness during the initialization process. Table~\ref{Table2} summarizes the four transfer learning methods used in the present study.

\begin{table}
  \footnotesize
  \caption{Summary of the transfer learning methods used in this study. ``$\alpha$'' in the PaPIR model represents $\alpha$ = $\sqrt{(2/({f_{i} + f_{o}})} \left( 1 - \lambda_2 \right)$.}
  \begin{minipage}{\linewidth}
    \begin{center}
    \begin{tabular}{l|l|l|l|l|l}
      \hline    
      Model   &  Description & Further & Objective & Initialization & Similar to\\ 
      &  & training &  & \\       
      \hline
      TL1 & Parameter sharing  & No & - & - & TL3 with $\lambda_1 = \infty$ \\ 
          &                    &    &   &   & PaPIR with $\lambda_1 = \infty$ \\       
      \hline      
      TL2 & Fine-tuning   & Yes & Target data misfit & \textbf{\textit{w}}$_{0}$ = \textbf{\textit{w}}$^{s}$ & TL3 with $\lambda_1 = 0$ \\
          &               &     &                    & \textbf{\textit{b}}$_{0}$ = \textbf{\textit{b}}$^{s}$ & PaPIR with $\lambda_1 = 0$, $\lambda_2 = 1$
      \\      
      \hline      
      TL3 & Parameter restriction  & Yes & Target data misfit  & \textbf{\textit{w}}$_{0}$ = \textbf{\textit{w}}$^{s}$  & PaPIR with $\lambda_2 = 1$ \\
          &                        &     & with $\lambda_1 \lVert \textbf{\textit{h}} - \textbf{\textit{h}}^s \rVert_2^2$ & \textbf{\textit{b}}$_{0}$ = \textbf{\textit{b}}$^{s}$  & \\      
      \hline     
      PaPIR & Parameter control  &  Yes &  Target data misfit &  \textbf{\textit{w}}$_{0}$ = & \\ 
            & via partial initialization &  & with $\lambda_1 \lVert \textbf{\textit{h}} - \textbf{\textit{h}}^s \rVert_2^2$ & \textit{N}( $\lambda_2$\textbf{\textit{w}}$^{s}$, $\alpha^2$) &  \\                                
            & and regularization &    &                     & \textit{b}$_{0}$ = \textit{N}($\lambda_2$\textbf{\textit{b}}$^s$, 0$^2$)  \\  
      \hline
    \end{tabular}
    \end{center}
  \end{minipage}
  \label{Table2}
\end{table}

\section{Results without transfer learning}\label{source_task}
The results of the PC-transport ROM for predicting the 0-D ignition process of hydrogen/air mixture over a wide range of $\phi$ at a given $T_0$ are presented. Transfer learning is not applied in this section. For the source task ($T_0 =$ 1000 K), a sufficient amount of training data is provided to train the ANN (i.e., $N_{\phi}$ = 30), and thus, the PC-transport ROM is expected to accurately capture the overall ignition characteristics of the hydrogen/air mixture. Subsequently, the effect of the number of training samples on the performance of the PC-transport ROM is investigated by gradually decreasing the number of training samples.

\subsection{Source task: $T_0 =$ 1000 K}\label{Source}
In the source task, the ANN models for predicting the reaction rate of PCs are trained by using the training dataset with $N_{\phi}$ of 30 ($\phi = 0.1 - 3.0$; $\Delta \phi$ = 0.1) at $T_0$ of 1000 K. A system of ODEs, Eq~\ref{PC-ODE}, is solved for 29 different 0-D simulations listed in the test dataset, and then the performance of the PC-transport ROM is evaluated against the  FOM by comparing $\tau_{\rm{ig}}$ between two different simulations. As mentioned earlier, $\tau_{\rm{ig}}$ is defined by the time at which the temperature gradient reaches its maximum value, and $\tau_{\rm{ig}}$ in the PC-transport ROM can be predicted after reconstructing the temperature profile from the results of PCs.

\begin{figure}
  \centerline{\includegraphics[width=70mm]{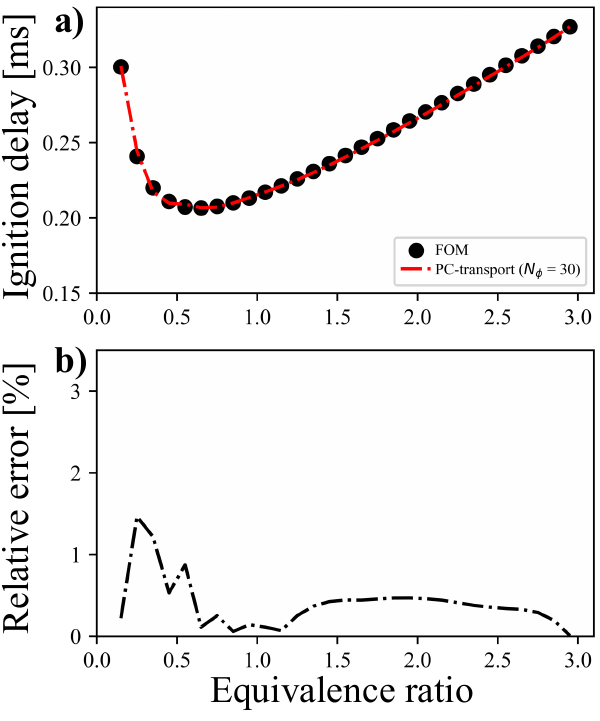}}
  \caption{Variations in (a) 0-D ignition delay time, $\tau_{\rm{ig}}$, predicted by FOM (solid symbol) and PC-transport ROM (dashed-dot line), and (b) the relative-error of the PC-transport ROM compared with FOM for the homogeneous hydrogen/air mixture with various $\phi$ (i.e., $\phi$ = 0.15 $\sim$ 2.95; $\Delta \phi$ =0.1) at $T_0$ = 1000 K.}\label{Fig4}
\end{figure}

Figure~\ref{Fig4} shows the variations in $\tau_{\rm{ig}}$ for the hydrogen/air mixture at $T_0$ = 1000 K with various $\phi$ predicted by the PC-transport ROM and FOM. As shown in the figure, $\tau_{\rm{ig}}$ predicted by the PC-transport ROM shows excellent agreement with the FOM. The relative percentage error is below 2 \% for the entire range of $\phi$, demonstrating that PC-transport ROM with a sufficient number of training samples can accurately reproduce the ignition process of a hydrogen/air mixture over a wide range of $\phi$ at a given $T_0$. Note that the relative percent error for the fuel-lean mixture is slightly higher than that for the fuel-rich mixture, which is attributed to the fact that the number of training samples assigned to the fuel-lean mixture is fewer than that assigned to the fuel-rich mixture. Also, $\tau_{\rm{ig}}$ shows a steeper variation with $\phi$ as $\phi$ becomes less than 0.5, which also affects the result.

\begin{figure}
  \centerline{\includegraphics[width=130mm]{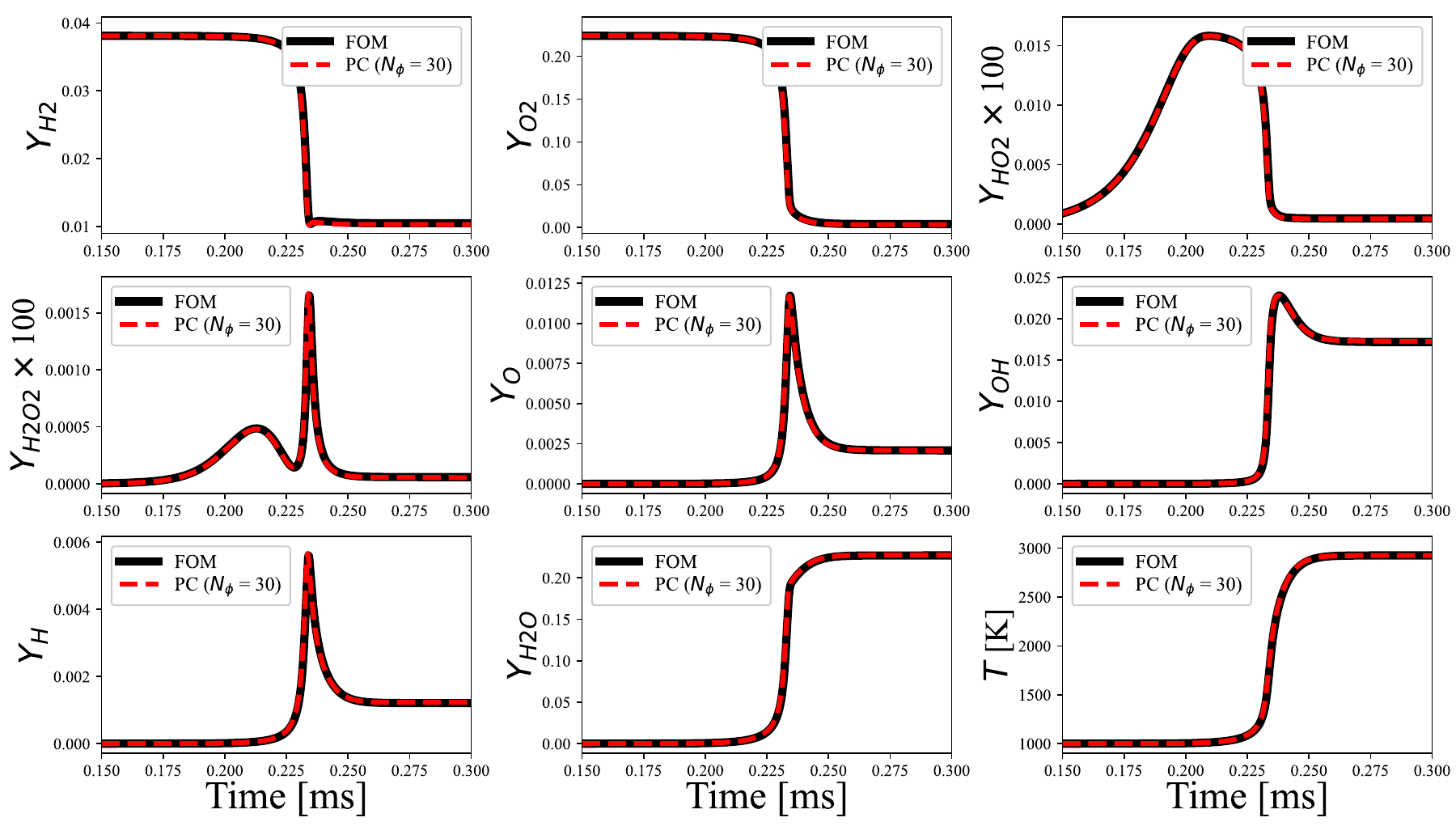}}
  \caption{Temporal evolution of the thermochemical state scalars of a homogeneous hydrogen/air mixture at $T_0$ = 1000 K and $\phi$ = 1.35. Solid line: FOM result, Dashed line: reconstructed from the PC-transport ROM result with $N_{\phi}$ = 30.}\label{Fig5}
\end{figure}

Figure~\ref{Fig5} presents the temporal evolution of the original thermochemical state vector of the hydrogen/air mixture at $T_0$ of 1000 K and $\phi$ of 1.35, as predicted by the PC-transport ROM and FOM. It is readily observed that the profiles of the thermochemical state variables reconstructed from the PC-transport ROM are in good agreement with the results from FOM for both major and minor species. This finding indicates that the number of PCs retained in this study ($N_{\rm{PC}} =$ 5) is sufficient to recover the original thermochemical state scalars, together with the successful validation of using the ANN model to convert from $\boldsymbol{\mathrm{\psi}}$ to $\boldsymbol{\mathrm{\theta}}$.

\subsection{Target task with data sparsity}\label{Target-Task}

The results of the PC-transport ROM for a target task where $T_0$ is 1050 K are investigated depending on the number of training samples. In this task, ANN models are trained by using each of the different training datasets, each with a different numbers of training samples (i.e., $N_{\phi}$ = 2 $-$ 30), and both \textit{a priori} and \textit{a posteriori} evaluations are carried out to assess the performance of the ROM depending on $N_{\phi}$. At a given $N_{\phi}$, the ANN model training is repeated 10 times to take into account the sensitivity of the model arising from the randomness of the initial parameters and the stochastic nature of the optimization process. The normalized root mean squared error (NRMSE) is adopted to \textit{a priori} quantify the error of the ANN for predicting $\dot{\boldsymbol{\mathrm{\omega}}}_{\boldsymbol{\mathrm{\psi}}}$ and is defined by 

\begin{equation}
  \textrm{NRMSE [\%]} =  \sqrt{\frac{\lVert \hat{\boldsymbol{\mathrm{\omega}}}_{\boldsymbol{\mathrm{\psi}},p} - \hat{\boldsymbol{\mathrm{\omega}}}_{\boldsymbol{\mathrm{\psi}}}\rVert_2^2}{\lVert \hat{\boldsymbol{\mathrm{\omega}}}_{\boldsymbol{\mathrm{\psi}}}\rVert_2^2}} \times 100 
  \label{NRMSE}
\end{equation}
where $\hat{\boldsymbol{\mathrm{\omega}}}_{\boldsymbol{\mathrm{\psi}},p}$ and $\hat{\boldsymbol{\mathrm{\omega}}}_{\boldsymbol{\mathrm{\psi}}}$ represent the normalized reaction rates of the PC vector predicted by the ANN model and obtained from the FOM, respectively. 

\begin{figure}
  \centerline{\includegraphics[width=70mm]{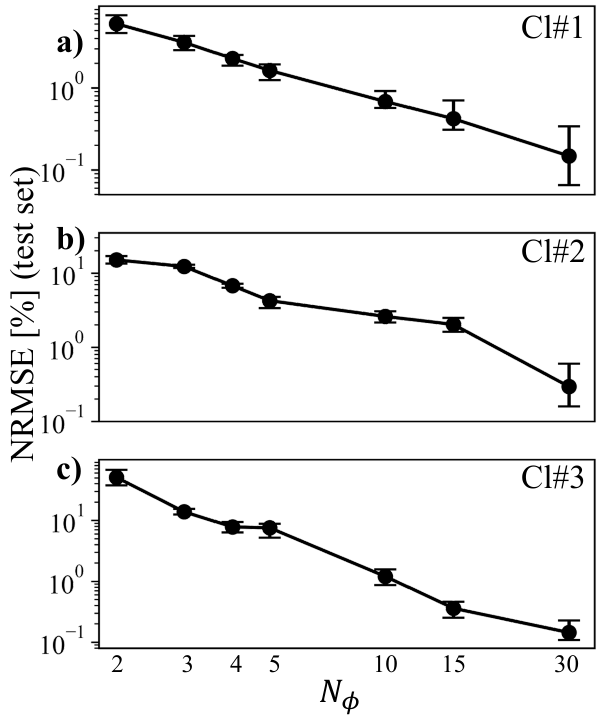}}
  \caption{Variations in NRMSE of the test set in the target task with $T_0$ of 1050 K as a function of $N_{\phi}$ for (a) Cluster 1, (b) Cluster 2, and (c) Cluster 3. The closed circle symbol represents the averaged NRMSE obtained from 10 repetitions of the ANN model training.}\label{Fig6}
\end{figure}

Figure~\ref{Fig6} shows the variations in NRMSE of the test set in the target task with $T_0$ of 1050 K and various $N_{\phi}$. The NRMSE of the test set generally shows a decreasing trend with an increase of $N_{\phi}$, such that the optimal value of NRMSE for the case with $N_{\phi} = 30$ approaches $O$(10$^{-1}$) [\%] for all the clusters. This outcome clearly indicates that the number of training samples plays a crucial role in determining the performance of the PC-transport ROM. In addition, the variations in the NRMSE as a result of repeating the ANN model training 10 times exhibit a noticeable fluctuation at a given $N_{\phi}$. Consequently, for Cl\#1 and Cl\#3, the worst cases with $N_{\phi} = 30$ have a similar magnitude of NRMSE compared to the best cases with $N_{\phi} = 15$. This result suggests that a multi-start based optimization algorithm would be necessary to obtain the nearest optimal neural network model at a given $N_{\phi}$. 

\begin{figure}
  \centerline{\includegraphics[width=70mm]{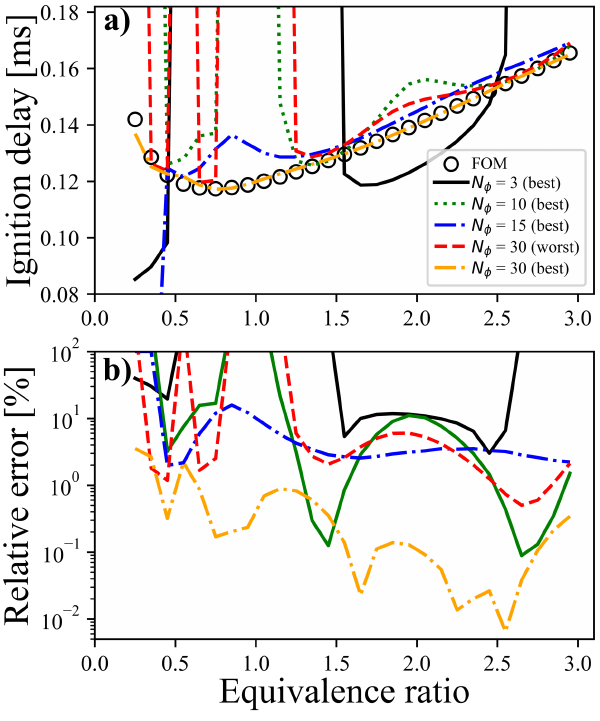}}
  \caption{Variations in (a) 0-D ignition delay time, $\tau_{\rm{ig}}$, predicted by the FOM (symbol) and PC-transport ROMs trained using a different number of training samples, and (b) the relative-error of the PC-transport ROMs compared with FOM for a homogeneous hydrogen/air mixture with various $\phi$ (i.e., $\phi$ = 0.15 $-$ 2.95; $\Delta \phi$ = 0.1) at $T_0$ = 1050 K.}\label{Fig7}
\end{figure}

Next, a series of 0-D simulations is carried out by using the ANN models trained with different numbers of training samples. For the case with $N_{\phi} = 30$, $\tau_{\rm{ig}}$ is predicted by using the best and worst ANN models out of  10 repetitions of the ANN model training. For the other cases, $\tau_{\rm{ig}}$ is predicted by using the best ANN model only. Figure~\ref{Fig7} summarizes the variations in $\tau_{\rm{ig}}$ for the hydrogen/air mixture with $T_0$ of 1050 K and various $\phi$ listed in the test set, predicted by using the ANN models with different $N_{\phi}$. 

Figure~\ref{Fig7} illustrates that the PC-transport ROM fails to capture the overall ignition characteristics of a hydrogen/air mixture with $N_{\phi} \leq$ 15. In this regard, the datasets with $N_{\phi} \leq 15$ are regarded as ``sparse datasets''. Note that the PC-transport ROM performs relatively well when the target equivalence ratio is adjacent to one of the equivalence ratios listed in the training dataset. For instance, the training dataset with $N_{\phi} = $ 3 consists of the 0-D simulation results with $\phi$ of 0.5 ,1.5, and 2.5, where the relative-error of the PC-transport ROM is relatively small near $\phi$ of 0.5, 1.5, and 2.5, while the performance of the PC-transport ROM declines as the target equivalence ratio moves farther from the training dataset. It is also important to note that even when a large amount of training samples are used (i.e., $N_{\phi} =$ 30), the simulation results occasionally do not agree well with the results from the FOM, consistent with the \textit{a priori} evaluation in Fig.~\ref{Fig6}. This result not only highlights that the number of training samples is a crucial part of optimizing the ANN model but also indicates that the uncertainty of the ANN model training is noticeable, and is primarily due to the stochastic nature of the training process and/or the randomness of the initial parameter.  

\section{Results with transfer learning}\label{TL}
In summary, the main issues associated with the PC-transport ROM for capturing the reaction rates of PCs with a sparse dataset are twofold: (1) the PC-transport ROM inaccurately predicts the reaction rate of the PCs over a wide range of $\phi$ due to a dearth of training samples, and (2) a multi-start-based optimization strategy is required to figure out the nearest optimal ANN model for a given training dataset. To address these issues, the previous knowledge gained from the source task ($T_0$ = 1000 K and $N_{\phi}$ = 30; see Sec.~\ref{Source}) is transferred to the target tasks. Four different transfer learning methods, TL1, TL2, TL3, and PaPIR, are applied to different target tasks in terms of the task similarity (i.e., $T_0$ difference between source and target task) and the degree of data sparsity in the target tasks (i.e., $N_{\phi}$ in the target task).

\subsection{TL1--TL3: General characteristics}\label{Target-Charact}

As a baseline case, the result of applying three transfer learning methods (TL1, TL2, and TL3) to the target task where $T_0$ = 1050 K and $N_{\phi} =$ 4 is discussed first. In this case, the difference of $T_0$ between source and target tasks is relatively small (i.e., $\Delta T$ = 50K), such that task similarity between the two tasks is considered to be high. The parameters obtained from the source task are used to train the ANN model in the target task in various ways.  

As shown in Table~\ref{Table2}, TL3 (parameter restriction) becomes equivalent to TL1 (parameter sharing) and TL2 (fine-tuning) as the value of the regularization parameter, $\lambda_1$, in Eq.~\ref{TL3} approaches near infinity and zero, respectively. Thus, the performance of the ANN model using three different transfer learning methods can be evaluated by adjusting the magnitude of $\lambda_1$. Similar to the previous cases where transfer learning is not employed, ANN training is repeated 10 times to evaluate the uncertainty of the ANN model training.

\begin{figure}
  \centerline{\includegraphics[width=130mm]{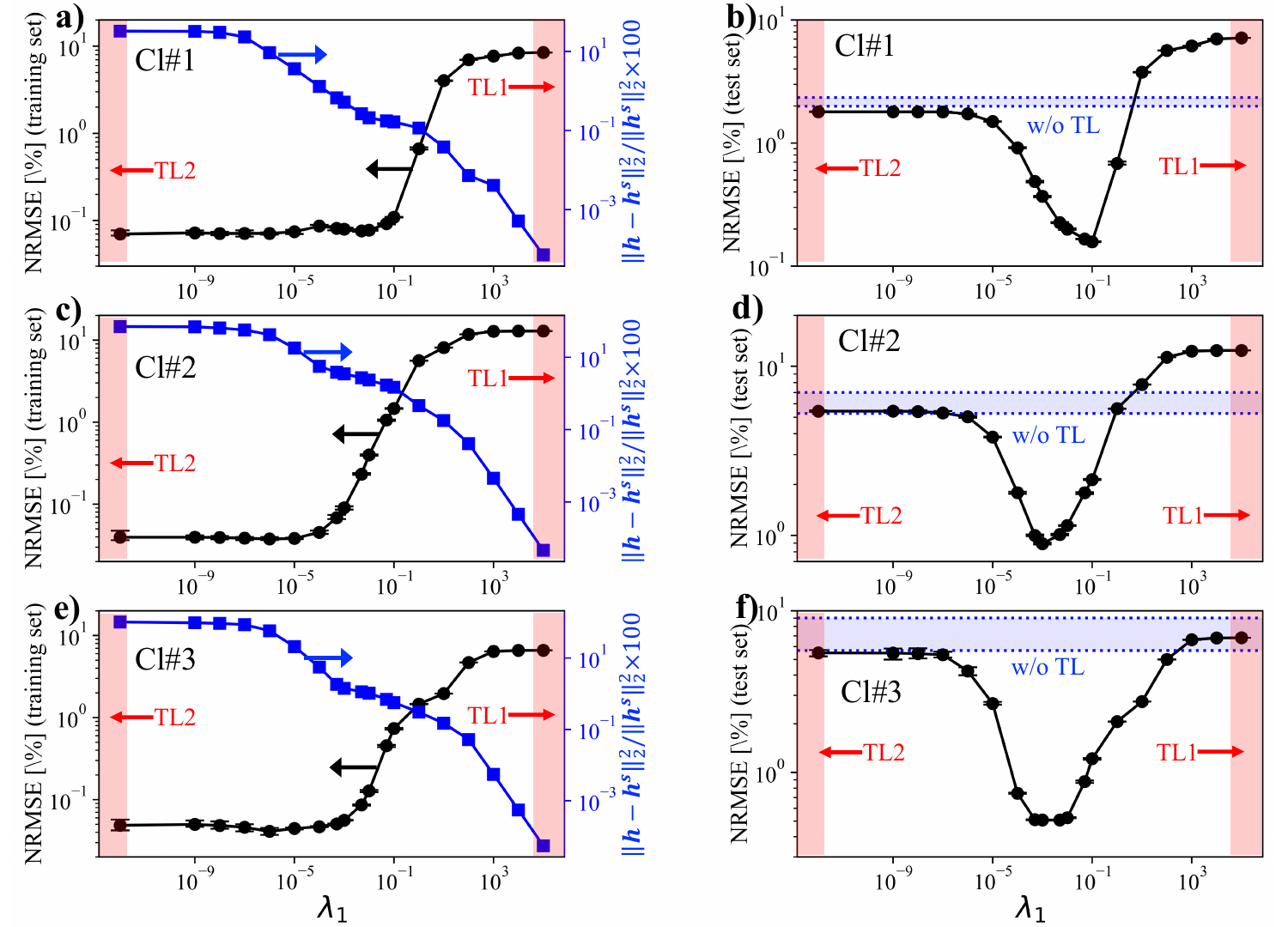}}
  \caption{\textit{{A} priori} evaluation of (left) the NRMSE for the training set and $\lVert \textbf{\textit{h}} - \textbf{\textit{h}}^{s} \rVert_2^2$/$\lVert \textbf{\textit{h}}^{s}  \rVert_2^2 \times 100$, and (right) the NRMSE for the test set for the target task with $T_0$ = 1050 K and $N_{\phi}$ = 4 as a function of $\lambda_1$. The highlighted regions on the right represent the range of NRMSE of the test set predicted by the PC transport model without applying transfer learning.}\label{Fig8}
\end{figure}

Figure~\ref{Fig8} shows the NRMSE values against the training and test sets, along with the percentage differences in the optimized parameters between the source and target tasks, represented by $\lVert \textbf{\textit{h}} - \textbf{\textit{h}}^{s} \rVert_2^2$/$\lVert \textbf{\textit{h}}^{s} \rVert_2^2 \times 100$, as a function of $\lambda_1$ for the case where $T_0$ = 1050 K and $N_{\phi}$ = 4. Several points are noted from the figure.

First, given that $\lambda_1$ serves as a penalty term during the fine-tuning, the NRMSE of the training set generally decreases with a decrease of $\lambda_1$ for all the clusters. On the other hand, $\lVert \textbf{\textit{h}} - \textbf{\textit{h}}^{s} \rVert_2^2$/$\lVert \textbf{\textit{h}}^{s} \rVert_2^2$ continues to increase as $\lambda_1$ decreases, indicating that the parameters in the target task become more dissimilar to those in the source task with a decrease of $\lambda_1$. This trend demonstrates that the magnitude of $\lambda_1$ mainly controls the degree of knowledge transfer from the source to the target task. 

Second, as the magnitude of $\lambda_1$ becomes sufficiently large (i.e., greater than 10$^{4}$), $\textbf{\textit{h}}$ becomes nearly identical to $\textbf{\textit{h}}^{s}$, illustrating that the transfer learning method for this case is equivalent to TL1, where the knowledge gained from the source task is fully transferred to the target task. In that case, the NRMSE of the test set can be notably higher than the case without applying transfer learning (see Figs.~\ref{Fig8}b and d as an example), indicating that TL1 would play a negative role in predicting the ignition delay of the hydrogen/air mixture for the target task even if the difference in $T_0$ is relatively small. This would be primarily attributed to the non-linear nature of chemical kinetics, where reaction rates are highly sensitive to temperature change.

Third, as the magnitude of $\lambda_1$ approaches near zero, the corresponding transfer learning method represents TL2, where the parameters obtained from the source task are used to initialize the parameters in the target task. In this scenario, $\lVert \textbf{\textit{h}} - \textbf{\textit{h}}^{s} \rVert_2^2$/$\lVert \textbf{\textit{h}}^{s} \rVert_2^2$ is relatively high, indicating that the knowledge acquired from the source task is prone to be lost during the fine-tuning process. Nonetheless, it is worth mentioning that the NRMSE of the test set using TL2 nearly equals that obtained from the best ANN model without applying transfer learning. Furthermore, the results of TL2 show less fluctuation from the 10 repetitions of training compared to the results without applying transfer learning. This suggests that initializing the ANN model of the target task with the parameters gained from the source task can be considered an appropriate initialization scheme, provided that the task similarity between the source and target task is high.

Lastly, the NRMSE of the test set reaches its minimum as the magnitude of $\lambda_1$ has a finite value (i.e., $\lambda_1 = $ $O$(10$^{-4}$ $-$ 10$^{-1}$)) for all the clusters. In this case, the resultant NRMSE is approximately an order of magnitude lower than that obtained from training the ANN model from scratch, clearly demonstrating that TL3 with an optimal value of $\lambda_1$ can remarkably improve the performance of the neural network model for the target task. Note that in order to achieve a comparable level of NRMSE as that obtained from TL3, the ANN model trained from scratch requires a larger number of $N_{\phi}$, ranging from 15 to 30, indicating that TL3 can reduce the requisite number of training samples up to eight times. Furthermore, the ANN model training with the use of the optimal value of $\lambda_1$ is nearly insensitive to the number of training repetitions. This is attributed to the well-known effect of  regularization on the stochastic optimization process. Hence, the overall number of training repetitions required to find the optimal ANN model in the target task can be significantly reduced by applying TL3 with an optimal value of $\lambda_1$, provided that the source and target tasks are similar in parameter space.

Based on the observations in Fig.~\ref{Fig8}, it can be inferred that the loss function in the target task is likely to contain multiple local minima such that the training result of the ANN model may not reach the global minima of the loss function, especially with the sparse dataset. One method to address this issue and improve the performance of the ANN model is to utilize the pre-trained ANN model. If the source and target tasks are similar to each other in the parameter space, then initializing the parameter in the target task ($\textbf{\textit{h}}_{0}$) with that from the source task ($\textbf{\textit{h}}^{s}$) can assist in searching for the optimal parameters. Hence, the performance of the ANN model with TL2 has the potential to show better performance compared to the ANN model trained from scratch. Furthermore, TL3 with the optimal value of $\lambda_1$ can help prevent $\textbf{\textit{h}}$ from significantly deviating from $\textbf{\textit{h}}^{s}$ during the fine-tuning step, which can enhance the accuracy in predicting $\tau_{\rm{ig}}$ in the target task. 

\begin{figure}
  \centerline{\includegraphics[width=70mm]{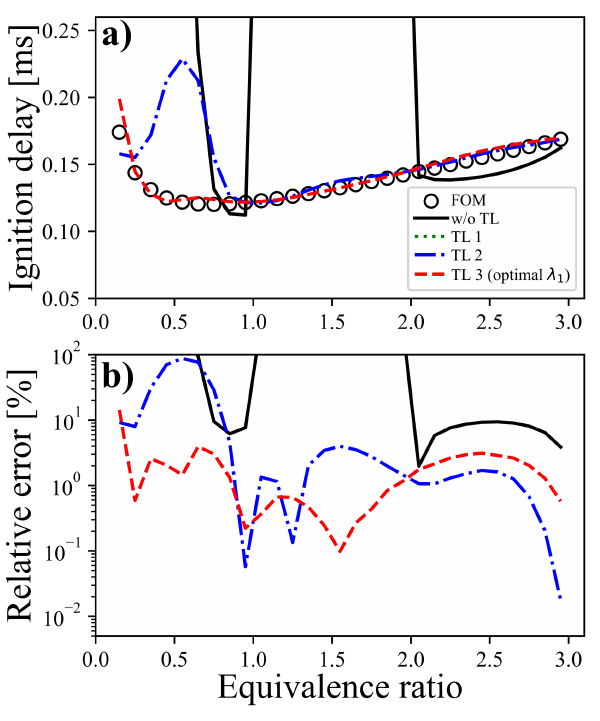}}
  \caption{Variations in (a) 0-D ignition delay time, $\tau_{\rm{ig}}$, predicted by FOM (solid symbol) and PC-transport ROMs trained by applying different transfer learning methods, and (b) the relative-error of the PC-transport ROMs compared with FOM for the homogeneous hydrogen/air mixture with various $\phi$ at $T_0$ = 1050 K. $N_{{\phi}}$ of the training set is set to 4.}\label{Fig10}
\end{figure}

Next, \textit{a posteriori} evaluation of the PC-transport ROM with different transfer learning strategies is carried out by performing a series of 0-D simulations of a hydrogen/air mixture using the PC-transport ROM for the target task, where $T_0$ = 1050 K and $N_{\phi}$ = 4. Figure~\ref{Fig10} shows the variations in $\tau_{\rm{ig}}$ as a function of $\phi$ depending on the different transfer learning methods. The PC-transport ROM fails to predict the overall ignition trend when the ANN is trained from scratch or trained with TL1. Note that the PC-transport ROM with TL1 inaccurately captures the early stage of 0-D ignition, leading to error accumulation over time. Consequently, the PC-transport ROM with TL1 fails to undergo a thermal runaway crossing for the full range of $\phi$, and hence, $\tau_{\rm{ig}}$ approaches infinity.

The overall performance of the PC-transport ROM with TL2 is better than the PC-transport ROM without applying the transfer learning method or with TL1, demonstrating the importance of the initialization scheme and the fine-tuning step on the result, respectively. Here, $\tau_{\rm{ig}}$ predicted by the PC-transport ROM with TL2 is in relatively good agreement with that from the FOM when $\phi$ is near those in the target task training dataset (i.e., $\phi$ of 0.2, 1.0, 2,0, and 3.0), while the performance of the ANN model decreases as $\phi$ becomes farther removed from those in the training dataset of the target task.

\begin{figure}
  \centerline{\includegraphics[width=160mm]{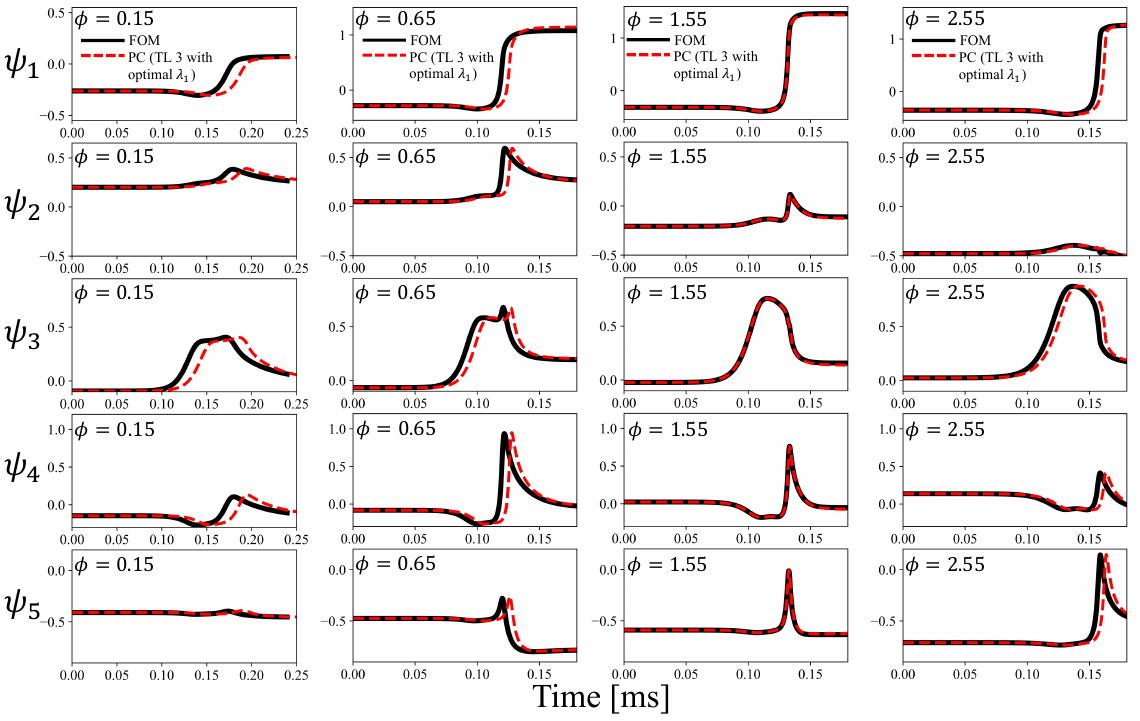}}
  \caption{Temporal evolution of the PCs that represent the homogeneous hydrogen/air mixture at $T_0$ = 1000 K and $\phi$ of 0.15, 0.65, 1.55, and 2.55, respectively (left to right). Solid line: PCs projected from the FOM result, Dashed line: PC-transport ROM using the optimal $\lambda_1$ in TL3.}\label{Fig11}
\end{figure}

For the PC-transport ROM with the optimal value of $\lambda_1$ (i.e., TL3), it is readily observed that the PC-transport ROM shows a good performance of predicting $\tau_{\rm{ig}}$ over a wide range of $\phi$. This result substantiates that the regularization-based transfer learning framework can increase the accuracy of the ANN model with the sparse training dataset. Figure~\ref{Fig11} presents the temporal evolution of the PCs with four different values of $\phi$ of 0.15, 0.65, 1.55, and 2.55, predicted by the FOM and the PC-transport ROM with an optimal value of $\lambda_1$. Although a slight time lag exists between the FOM and the PC-transport ROM results due to data sparsity, the PC-transport ROM can reasonably capture the onset of ignition and the subsequent equilibrium period of the PCs. 

\subsection{TL1--TL3: Parametric study in terms of task similarity and data sparsity}\label{Param}
\begin{figure}
  \centerline{\includegraphics[width=130mm]{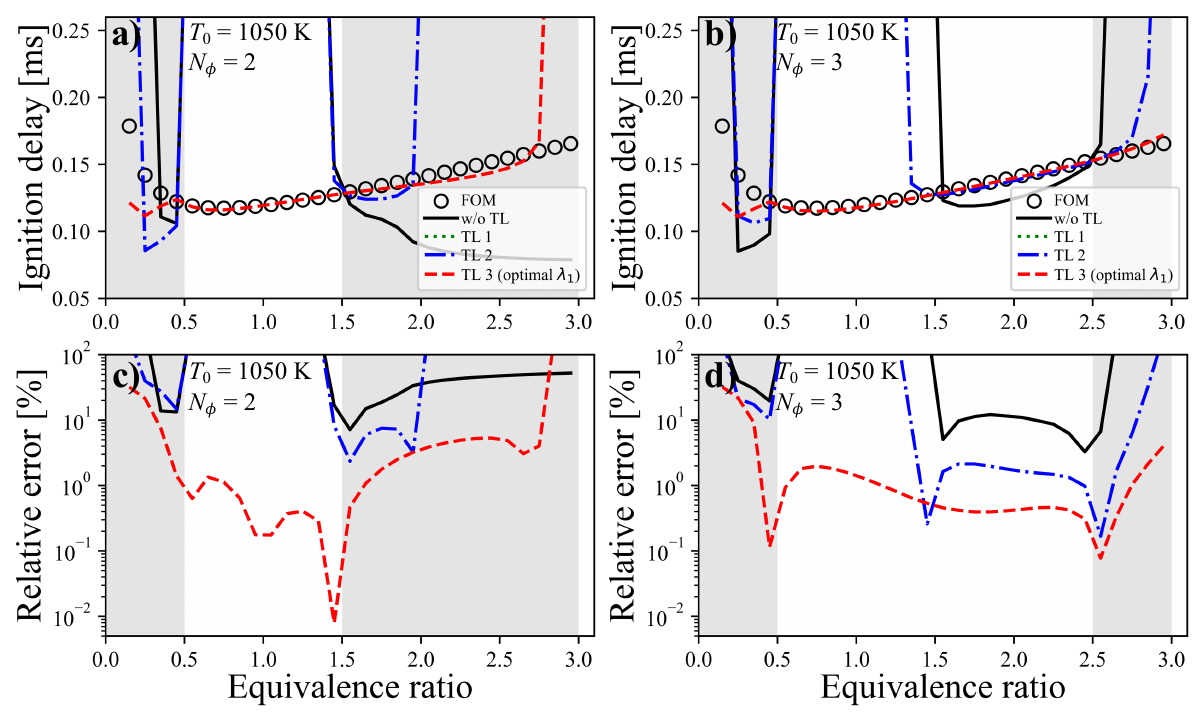}}
  \caption{ Variations in (top) $\tau_{\rm{ig}}$ predicted by FOM (solid symbol) and PC-transport ROMs trained by applying different transfer learning methods, and (bottom) the relative-error of the PC-transport ROMs compared with FOM for the homogeneous hydrogen/air mixture with various $\phi$ at $T_0$ = 1050 K. Here, $N_{{\phi}}$ of the training set is (left) 2, and (right) 3, respectively. The highlighted region represents the cases where $\phi$ is out of the range of the training dataset in the target task.}\label{Fig13}
\end{figure}

Additional parametric studies are carried out by varying $N_{\phi}$ or increasing $T_0$ to 1300 and 1400 K. Figure~\ref{Fig13} shows the variations in $\tau_{\rm{ig}}$ as a function of $\phi$ for the target task with $T_0$ of 1050 K and decreasing $N_{\phi}$ to 2 and 3 by applying different transfer learning methods. Note that as $N_{\phi}$ decreases to 2 and 3, the training dataset is intended not to cover the entire range of the test dataset (i.e., $\phi$ = 0.5 and 1.5 for $N_{\phi}$ = 2, and $\phi$ = 0.5, 1.5, and 2.5 for $N_{\phi}$ = 3) such that there exist several test cases where $\phi$ is outside of the range of the training dataset (see the highlighted regions in Fig.~\ref{Fig13}). The overall variations in $\tau_{\rm{ig}}$ predicted by applying different transfer learning methods show a similar trend regardless of the change of $N_{\phi}$. The PC-transport ROM with TL1 fails to capture the onset of ignition of the hydrogen/air mixture for the entire range of $\phi$, while the results with TL2 show a better performance than those without applying transfer learning. The PC-transport ROM with the optimal value of $\lambda_1$ in TL3 outperforms all the other models. 

As expected, the accuracy of the PC-transport ROM notably decreases as the target $\phi$ of the 0-D simulation is outside of the range of the training dataset, which is a well-known drawback of machine learning models for extrapolation. Nonetheless, the result of the PC-transport ROM with TL3 shows a relatively-good performance even for the cases where $\phi$ is outside of the range of the training set. This result implies that the previous knowledge obtained from the source task helps increase the accuracy of the extrapolation of the ANN model, consistent with previous findings \cite{Humbird20}. 

\begin{figure}
  \centerline{\includegraphics[width=130mm]{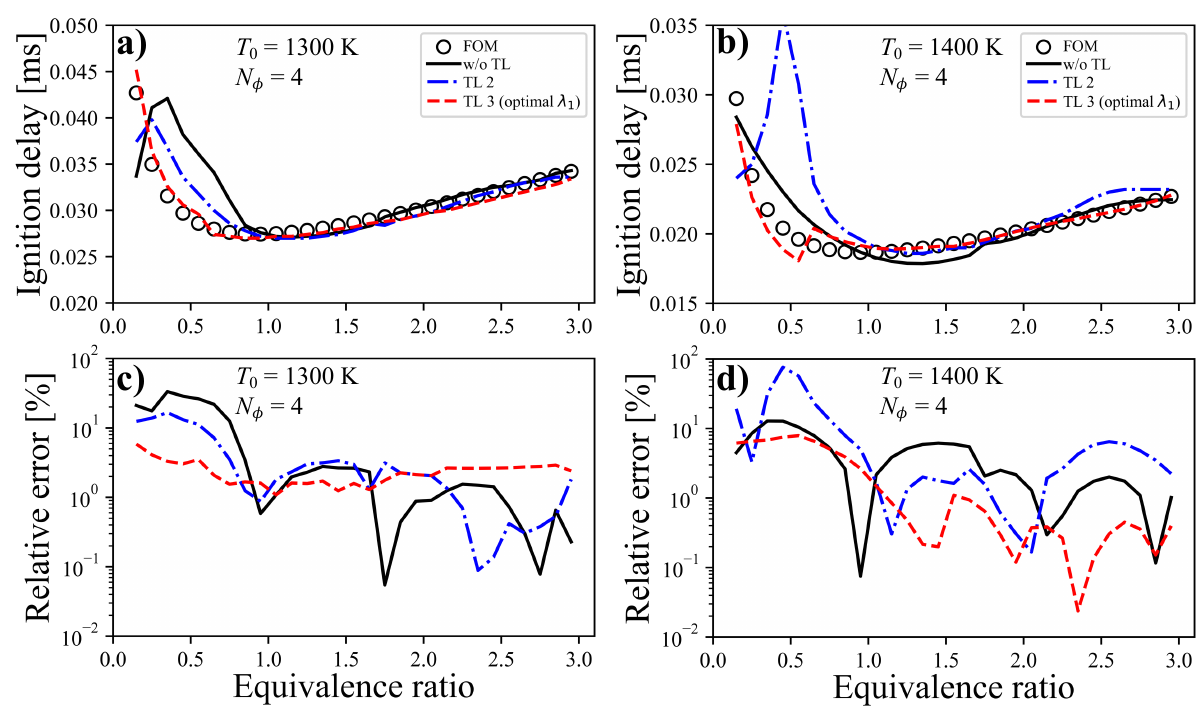}}
  \caption{Variations in (top) $\tau_{\rm{ig}}$ predicted by (solid symbol) FOM and PC-transport ROMs trained by applying different transfer learning methods, and (bottom) the relative-error of the PC-transport ROMs compared with FOM for the homogeneous hydrogen/air mixture with various $\phi$ at (left) $T_0$ = 1300 K and (right) $T_0$ = 1400 K.}\label{Fig16}
\end{figure}

Next, target tasks are considered where $T_0$ is increased further (i.e., $T_0$ = 1300, and 1400 K) such that the task similarity between source and target tasks decreases. Figure~\ref{Fig16} shows the variations in $\tau_{\rm{ig}}$ for the hydrogen/air mixture with $T_0$ of 1300 and 1400 K and $N_{\phi}$ of 4 by using various ANN models with or without applying transfer learning methods. For the cases with $T_0$ = 1300 K, it is found that the PC-transport ROM with the optimal value of $\lambda_1$ in TL3 shows a reasonable performance over the entire range of $\phi$, while the results without applying transfer learning or with applying TL2 exhibit a noticeable error in predicting  ignition of a lean mixture. As $T_0$ in the target task further increases to 1400 K, on the other hand, the result with TL3 shows only a marginal improvement compared to the other cases. This result demonstrates that the performance of the regularization-based transfer learning method decreases with a decrease of task similarity between source and target tasks.

\subsection{PaPIR: unified transfer learning method}\label{Target-PaPIR}
Lastly, the performance of the unified transfer learning method, PaPIR, in the different target tasks is investigated. As discussed in Section~\ref{TL-methods}, the central idea of PaPIR is to control the degree of knowledge transfer from the source to target task by adjusting the magnitudes of $\lambda_1$ and $\lambda_2$, which are associated with the regularization and initialization of the ANN model in the target task, respectively. Unlike TL3, \textbf{\textit{h}}$_0$ in PaPIR can be distributed by either a normal distribution function following the Xavier normal initialization method ($\lambda_2 = 0$), or \textbf{\textit{h}}$^2$ ($\lambda_2 = 1.0$), or inbetween the two ($0 < \lambda_2 < 1$). In this regard, the effect of the initialization on the performance of transfer learning can be investigated by varying $\lambda_2$ in PaPIR.  

\begin{figure}
  \centerline{\includegraphics[width=160mm]{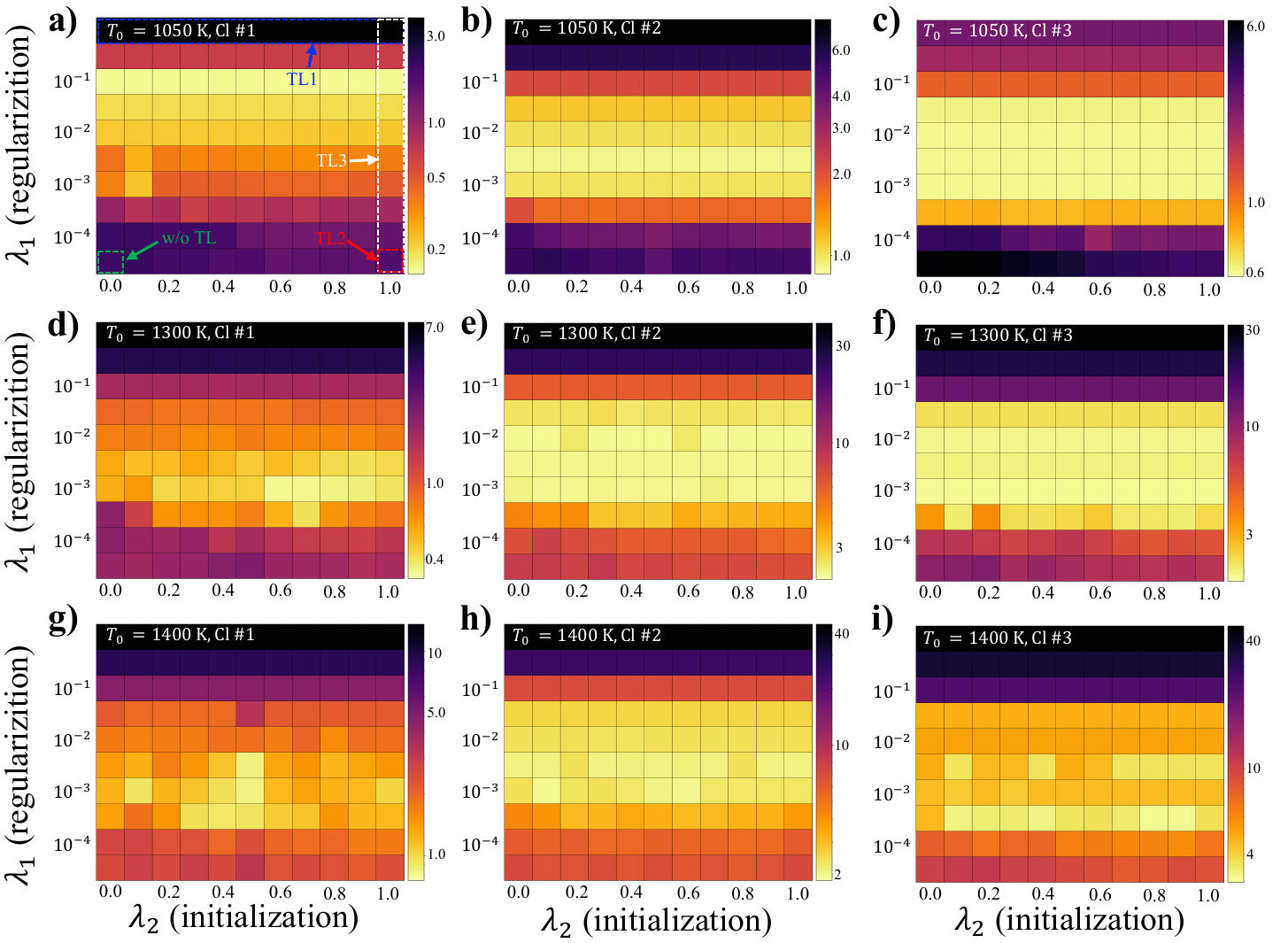}}
  \caption{Distributions of the best achievable value of NRMSE [\%] using PaPIR as a function of $\lambda_1$ and $\lambda_2$ for the different test datasets out of 10 repetitions of the ANN model training. The target task is varied ranging from $T_0 =$ 1050, 1300, and 1400 K (top to bottom) for Cluster 1, 2, and 3 (left to right) with $N_{\phi}$ of 4.}\label{Fig17}
\end{figure}

Figure~\ref{Fig17} presents the \textit{{a} priori} result of the best achievable NRMSE of $\dot{\boldsymbol{\mathrm{\omega}}}_{\boldsymbol{\mathrm{\psi}}}$ of the test set for three different target tasks (i.e., $T_0$ = 1050, 1300, and 1400 K with  $N_{\phi}$ of 4) as a function of $\lambda_1$ and $\lambda_2$, conditional on each of the three different clusters. Consistent with the previous results, the best achievable (or minimum) value of NRSME is evaluated by repeating the ANN model training 10 times at a given $\lambda_1$ and $\lambda_2$. As shown in Fig.~\ref{Fig17}a, PaPIR covers all the transfer learning methods discussed in the present study, namely, TL1, TL2, and TL3. In general, the results with TL1 exhibit a large error and increase with an increase of $T_0$ in the target task. In TL3, there exists an optimal value of $\lambda_1$ that results in a lower NRMSE than for the results either without applying transfer learning or with TL2. 

 
For the case where task similarity is relatively high (i.e., $T_0 = 1050$ K in the target task; see Figs.~\ref{Fig17}a--c), the value of NRMSE is mainly governed by the regularization parameter $\lambda_1$, whereas it is largely unaffected by a change of the initialization parameter, $\lambda_2$. Since the source and target tasks are similar to each other in this case, the optimal value of $\lambda_1$ is relatively large (e.g.,  $\lambda_1$ = 10$^{-1}$ in Cl\#1). Given that a regularization term serves to convexify the objective function, a relatively large magnitude of $\lambda_1$ leads the ANN model to be insensitive to a change in the initialization scheme. Consequently, PaPIR does not outperform TL3 when the task similarity between source and target tasks is high. The best achievable values of the NRMSE depending on the different transfer learning methods are summarized in Table~\ref{Table3}.

As $T_0$ in the target task increases to 1300 K, results with Cl\#1 (Fig.~\ref{Fig17}d) show that the NRMSE of the test dataset attains its minimum at a relatively low magnitude of $\lambda_1$ (= 10$^{-3}$). Although the overall variations of the NRMSE are still mainly governed by $\lambda_1$, the NRMSE is no longer invariant to a change of $\lambda_2$ at the optimal value of $\lambda_1$, indicating that $\lambda_2$ starts to play a role in the optimization of the ANN model. Since the magnitude of the optimal value of $\lambda_1$ decreases compared to the case with $T_0 = 1050$ K, the complexity of the loss function at the optimal value of $\lambda_1$ increases, and consequently, the training result can be varied with the different initialization schemes. This finding indicates that an initialization scheme becomes important in the framework of transfer learning as the task similarity between the source and target task becomes relatively low. Note that in Fig.~\ref{Fig17}d, the ANN model exhibits a slightly-lower magnitude of NRMSE at $\lambda_2$ of 0.7 compared to that of 1.0, illustrating the potential advantage of PaPIR over TL3. Readers are referred to Table~\ref{Table3} to quantify the difference of NRMSE between PaPIR and TL3. For the results of Cl\#2 and Cl\#3, on the other hand, $\lambda_2$ still plays a major role in determining the best achievable value of NRMSE (see Figs.~\ref{Fig17}e and f). This would be because the decrease of task similarity for these clusters is not as pronounced as for Cl\#1.

As $T_0$ in the target task further increases to 1400 K, it is found that $\lambda_1$ still shows a dominant effect on the NRSME compared to $\lambda_2$, demonstrating that the primary factor of determining the performance of transfer learning is a regularization parameter in general. Nonetheless, there are several cases where the ANN model trained with $\lambda_2 < 1$ exhibits a lower magnitude of NRMSE compared to the best candidate obtained from TL3 (see Fig.~\ref{Fig17}g). Note that at $T_0$ = 1400 K, the ratio of the best achievable NRMSE obtained from PaPIR to TL3 is 0.827, 0.984, and 0.866, for Cl\#1, Cl\#2, and Cl\#3, respectively. This result shows that adjusting the initial values of the parameters in the target task can further enhance the performance of transfer learning in the target task with a sparse dataset, especially when the task similarity between source and target tasks is low such that the optimal value of the regularization parameter, $\lambda_1$, becomes relatively low. 

\begin{table}[h]
\centering
        \begin{tabular}{c|c|c|c|c|c|c|c|c|c}
            \hline
            \multirow{2}{*}{Model} & \multicolumn{3}{c|}{$T_0$ = 1050 K} & \multicolumn{3}{c|}{$T_0$ = 1300 K} & \multicolumn{3}{c}{$T_0$ = 1400 K} \\\cline{2-10}
             & C1 & C2  & C3 & C1 & C2  & C3 & C1 & C2  & C3 \\\cline{1-10}    
            w/o TL & 1.9818 & 5.4682 & 6.0114 & 1.7982 & 8.6369 & 10.961 & 2.2756 & 6.8892 & 11.000 \\             \hline
            TL1 & 3.7541 & 7.7887 & 2.7417 & 7.3993 & 38.760 & 31.030 & 13.134 & 44.246 & 46.182 \\                \hline
            TL2 & 1.7861 & 5.3703 & 5.3862 & 1.9408 & 6.9736 & 8.6645 & 2.1617 & 6.9680 & 11.038 \\                \hline
            TL3 & \textbf{0.1570} & \textbf{0.8949} & \textbf{0.5108} & 0.3893 & \textbf{2.2277} & \textbf{1.8973} & 1.0157 & 2.1242 & 3.6054 \\                \hline
            PaPIR & \textbf{0.1570} & \textbf{0.8967} & \textbf{0.5105} & \textbf{0.3316} & \textbf{2.2227} & \textbf{1.8963} & \textbf{0.8401} & \textbf{2.0896} & \textbf{3.1229} \\                \hline

        \end{tabular}
    \caption{Best achievable value of NRMSE [\%] by using different transfer learning methods for the test dataset with various $T_0$ and $N_{\rm{\phi}} = 4$, out of 10 repetitions of ANN model training.}
    \label{Table3}
\end{table}

\begin{figure}
  \centerline{\includegraphics[width=70mm]{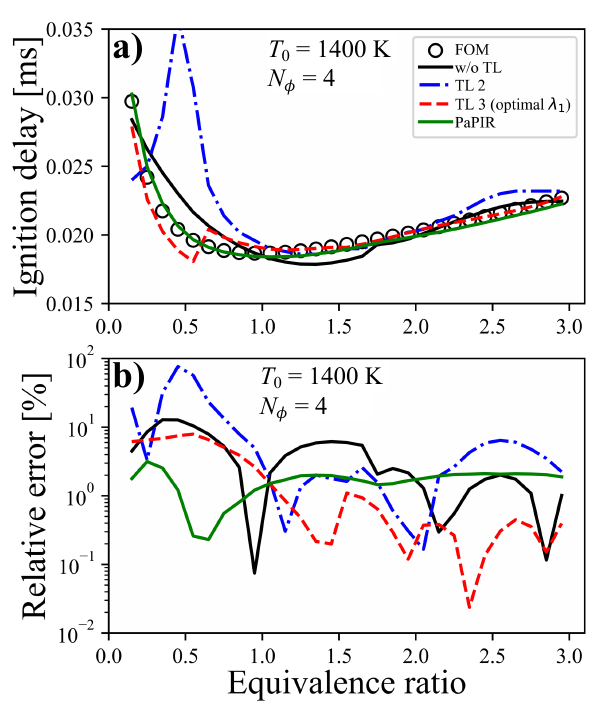}}
  \caption{Variations in (top) $\tau_{\rm{ig}}$ predicted by the (open symbol) FOM and PC-transport ROMs trained by applying different transfer learning methods, and (b) the relative-error of the PC-transport ROMs compared with the FOM for a homogeneous hydrogen/air mixture with various $\phi$ at $T_0$ = 1400 K and $N_{\phi}$ = 4.}\label{Fig18}
\end{figure}

To further investigate the advantage of PaPIR over other transfer learning methods, especially when the task similarity is relatively low, Figure~\ref{Fig18} presents the variations in $\tau_{\rm{ig}}$ for a hydrogen/air mixture with $T_0$ of 1400 K and $N_{\phi}$ of 4, predicted by the FOM and the PC-transport ROMs with different transfer learning methods. This figure clearly shows that $\tau_{\rm{ig}}$ predicted by PaPIR shows excellent agreement with that from the FOM over the entire range of $\phi$, which is clearly distinct from the other models. Although the relative-error obtained from PaPIR is slightly higher than that from TL3 or from the PC-transport ROM without applying transfer learning at $\phi > 1$, the PC-transport ROM with PaPIR shows a more robust performance for predicting the oxidation process of hydrogen/air mixture over a wide range of $\phi$. 

One may argue that under the data-sparse scenario, the number of test datasets is also likely to be insufficient, rendering it  infeasible to find the optimal values of $\lambda_1$ and $\lambda_2$ by relying on the test dataset. As a future work, a systematic way of estimating the optimal values of those two parameters without relying on a test dataset will be investigated. One practical example would be adopting the L-curve criterion, a well-known heuristic method to find the optimal regularization parameter without relying on the test dataset \cite{Hansen92}. Also as future work, we will further investigate different methods to ``partially'' transfer the knowledge of the pre-trained neural network model in the target task, such as applying Bayesian transfer learning methods.

\section{Conclusions}
In this study, various transfer learning methods were applied to the prediction of the reaction rate of the PCA-based low-dimensional manifold that represents the ignition process of a homogeneous hydrogen/air mixture in a constant volume reactor. A sufficient number of training samples spanning a wide range of $\phi$ was provided in the source task where $T_0 = 1000$ K, whereas the number of training datasets was assumed to be sparse in the target task where $T_0 > 1000$ K. The effect of the number of training samples on the performance of the PC-transport ROM was first investigated, followed by the application of three different transfer learning approaches (i.e., TL1, TL2, and TL3) to the different target tasks. To this end, a unified transfer learning framework was proposed in this study to elucidate the role of initialization and regularization on the performance of transfer learning. The following results are highlighted from the present study: 

\begin{itemize}
    \item In general, the number of training datasets played a primary role in determining the performance of the model. Without applying transfer learning, the PC-transport ROM failed to reproduce the ignition process of a hydrogen/air mixture with a sparse dataset (i.e., $N_{\phi} \leq 15$). It was also found that the PC-transport ROM without transfer learning shows a relatively-good accuracy for the test cases when the initial condition of the ROM is adjacent to that included in the training dataset.    
    \item Three different transfer learning methods, parameter sharing (TL1), fine-tuning (TL2), and parameter restriction (TL3), were then applied to the target task where $T_0 = 1050$ K and $N_{\phi}$ = 4. The PC-transport ROM using TL1 led to a significant error in predicting the reaction rate of PCs, while the PC-transport ROM with TL2 showed a slightly better performance than that without applying transfer learning approaches. An optimal value of the regularization parameter $\lambda_1$ in TL3 led to a remarkable decrease in the NRMSE of the test dataset. It was also illustrated that the profiles of the 0-D ignition delay predicted by the PC-transport ROM with TL3 exhibit good agreement with those obtained from the FOM, demonstrating the importance of the regularization-based transfer learning method.
    \item Parametric studies were performed by varying $T_0$ and $N_{\phi}$ in the target tasks to investigate the effect of task similarity and data sparsity in the target task on the performance of the different transfer learning methods, respectively. It was found that the knowledge from the source task helped predict the ignition process of a hydrogen/air mixture outside of the $\phi$ range in the training dataset, demonstrating the advantage of applying transfer learning for extrapolation. As $T_0$ in the target task was increased to 1400 K, the performance of TL3 is no longer remarkable due to the decrease of the task similarity between the source and target task.
    \item A novel transfer learning approach, PaPIR, was applied to the various target tasks. When the task similarity between the source and target tasks is high, the effect of the initialization parameter, $\lambda_2$, has a negligible effect on the NRMSE of the test set of the target task, while the minimum of the NRMSE is primarily determined by $\lambda_1$. The optimal value of $\lambda_1$ decreased with a decrease of task similarity, such that the effect of different initialization schemes on the result became noticeable. Although $\lambda_1$ still had a dominant effect on the result, an additional performance improvement could be achieved by changing the magnitude of $\lambda_2$, illustrating the potential advantage of PaPIR. 
\end{itemize}

\section*{Declaration of Competing Interest}\label{Competing}
The authors declare that they have no known competing financial interests or personal relationships that could have appeared to influence the work reported in this paper.

\section*{Acknowledgements}\label{Acknowledgements}
This work was supported by the Laboratory Directed Research and Development program at Sandia
National Laboratories (Project 222361), a multimission laboratory managed and operated by National Technology and
Engineering Solutions of Sandia LLC, a wholly owned subsidiary of Honeywell International Inc. for the
U.S. Department of Energy’s National Nuclear Security Administration under contract DE-NA0003525.
This report describes objective technical results and analysis. Any subjective views or opinions that might
be expressed in the report do not necessarily represent the views of the U.S. Department of Energy or the
United States Government.

The authors would like to acknowledge the contributions of Anuj Kumar at North
Carolina State University for sharing the code for the PC-transport ROM.


\bibliographystyle{elsarticle-num}
\bibliography{Bibliography}

\end{document}